\documentclass{article}
\usepackage{multirow}
\usepackage{array}
\usepackage{spconf,amsmath,graphicx}
\usepackage{amsfonts,amssymb}
\usepackage{hyperref} %**IVM

\def\x{{\mathbf x}}
\def\y{{\mathbf y}}

\def\z{{\mathbf z}}
\def\w{{\mathbf w}}
\def\B{{\mathbf B}}

\title{Joint generative-contrastive representation learning for anomalous sound detection}
\name{Xiao-Min Zeng$^1$,Yan Song$^1$,Zhu Zhuo$^3$,Yu Zhou$^3$,Yu-Hong Li$^3$,Hui Xue$^3$,Li-Rong Dai$^1$,Ian McLoughlin$^{1,2}$}
\address{$^1$National Engineering Research Center of Speech and Language Information Processing, \\
	University of Science and Technology of China, Hefei, China. \\
	$^2$ICT Cluster, Singapore Institute of Technology, Singapore. \\
	$^3$Alibaba Group, China.
	\thanks{Yan Song is the corresponding author. This work was supported by the Leading Plan of CAS~(XDC08030200) and The National Kay Research and Development Program of China under Grant 2020AAA0107705.}
}
\begin{document}
%	\ninept
	%
	\maketitle
	\begin{abstract}
		In this paper, we propose a joint generative and contrastive representation learning method~(GeCo) for anomalous sound detection~(ASD).
		GeCo exploits a Predictive AutoEncoder (PAE) equipped with self-attention as a generative model to perform frame-level prediction.
		The output of the PAE together with original normal samples, are used for supervised contrastive representative learning in a multi-task framework.
		Besides cross-entropy loss between classes, contrastive loss is used to separate PAE output and original samples within each class.
		GeCo aims to better capture context information among frames, thanks to the self-attention mechanism for PAE model. 
		Furthermore, GeCo combines generative and contrastive learning from which we aim to yield more effective and informative representations, compared to existing methods.
		Extensive experiments have been conducted on the DCASE2020 Task2 development dataset, showing that GeCo outperforms state-of-the-art generative and discriminative methods.
	\end{abstract}

	\begin{keywords}
		predictive autoencoder, contrastive learning, representation learning, anomalous sound detection
	\end{keywords}
	\section{Introduction}
	\label{sec:intro}
	\begin{figure*}[t]
		\centering
		\includegraphics[width=0.985\linewidth]{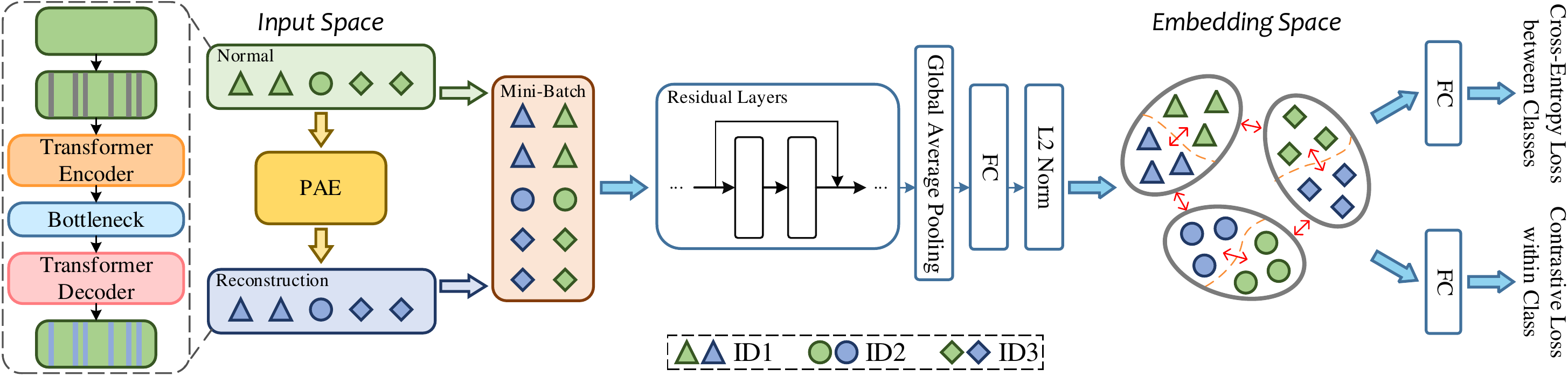}
		\caption{An overview of the proposed joint generative-contrastive~(GeCo) training framework.
			Predictive AutoEncoders are pretrained for each machine type and then utilized to generate negative samples for contrastive learning.
			Together with normal samples, reconstruction samples generated by PAE are used to train the feature extractor.
			The model not only distinguishes machine IDs but also determines whether the input is a reconstruction sample.}
		\label{fig1}
	\end{figure*}
	The task of analysing sound recordings to automatically detect whether a given clip is abnormal or not is referred to as anomalous sound detection~(ASD).
	It has wide application in areas such as automatic surveillance and machine condition monitoring, and has received increasing research attention thanks to DCASE\footnote{DCASE: Detection and Classification of Acoustic Scenes and Events, \url{https://dcase.community}.} challenges.
	The task is significantly complicated by the difficulty of collecting and labeling anomalous sounds~\cite{Koizumi2020} which often occur sporadically.
	
	Existing systems from recent DCASE challenges that have achieved promising performance, mainly focus on describing the distribution of normal data using machine learning methods, for sounds emitted from target machines~\cite{Koizumi2020}.
	Conventional ASD systems exploit generative models~\cite{Koizumi2020,9576445,9414662}, which detect anomalies based on the reconstruction error learned from the normal sounds.
	In 2020, Interpolation DNN~(IDNN) was proposed~\cite{suefusa2020anomalous}, where a model is trained to predict the removed centre frame of a spectrogram from the surrounding frames, leading to improved frame-wise representation, especially for non-stationary sounds.
	ID-aware autoencoders can also utilize class information (\textit{e.g.}, machine class or attribute) to assist the model in avoiding confusion from anomalous or normal sounds in different classes~\cite{Hayashi2020}.
	Many discriminative CNN-based methods (e.g. ResNet~\cite{hojjati2022self, chen2022self}, MobileNetV2~\cite{giri2020self}, and STgram~\cite{liu2022anomalous}), have also been applied to the ASD task.
	Such methods mainly focus on learning representations from the perspective of outlier exposure~(OE)~\cite{hendrycks2018deep} or self-supervised learning~\cite{chen2022self}, where sounds from different machines are considered to be pseudo anomalies. 
	ASD performance can be further improved by fusing anomaly scores from generative and discriminative systems~\cite{giri2020self}.
	
	In general, the key to improving ASD performance is to learn a compact and discriminative feature representation from normal sound data.
	However, in AE-based methods~\cite{Koizumi2020, suefusa2020anomalous}, multiple frames of a spectrogram are reshaped to a one-dimension input feature, which fails to take full advantage of temporal context information.
	Wichern et al.~\cite{wichern2021anomalous} proposed attentive neural processes (ANP) to improve performance by introducing an attention mechanism between coordinates and magnitude.
	Meanwhile, discriminative learning methods~\cite{chen2022self, giri2020self, liu2022anomalous} generally learn feature representations to discriminate known classes in the ideal case~\cite{perera2020generative}.
	However, since actual anomalous sounds occur only sporadically and are naturally highly diverse, a learned feature representation is unlikely to be sufficient for detecting unseen anomalies having a distribution different from that of seen data.
	
	In this paper, we propose a joint generative-contrastive method~(GeCo) to learn more effective representations, as shown in Fig.~\ref{fig1}.
	A Predictive AutoEncoder~(PAE) consisting of transformer blocks, is proposed to employ a masking and prediction strategy to learn a frame-wise representation.
	Its self-attention mechanism is exploited to model relationships across neighbouring frames as in~\cite{vaswani2017attention, he2022masked}.
	PAE-reconstructed samples, together with original normal data, are used for joint generative and contrastive learning under the multi-task framework.
	Cross-Entropy~(CE) loss is used to discriminate known classes while supervised contrastive loss helps to discriminate the reconstruction and original samples within a class.
	GeCo aims to build a model which learns a more informative representation with the help of PAE.
	The supervised contrastive loss can further enhance the compactness and discrimination of the learned representations. 
	
	We evaluate the effectiveness of the GeCo learned representation on the DCASE2020 Task2 development dataset, revealing a significant improvement over the state-of-the-art systems.
	GeCo achieves 93.97\% AUC and 87.34\% pAUC comfortably outperforming the 90.47\% AUC and 83.61\% pAUC of the DCASE2020 winning system.
	
	% 实验验证
	\section{Method}
	\label{sec:method}
	In this section, we will introduce the proposed GeCo in detail, including the generative model PAE, and joint generative and contrastive learning procedure employing a multi-task framework, as shown in Fig.~\ref{fig1}.

	\subsection{Predictive AutoEncoder~(PAE)}	\label{section2-1}
	The Predictive AutoEncoder~(PAE) uses an encoder-decoder architecture similar to~\cite{suefusa2020anomalous}, to predict unseen frames based on the self-attention mechanism~\cite{vaswani2017attention, he2022masked}.  
	Given an acoustic feature $\x$ with $n$ frames, input feature $\x_{m}$ is obtained by randomly masking part of the frames of $\x$.
	An encoder $E$ is trained to learn latent representations from the input sequence, while a decoder $D$ is trained to predict the masked features from latent representations.
	Both encoder $E$ and decoder $D$ comprise several transformer blocks.
	Due to the introduction of a self-attention mechanism, context information is obtained across frames, which is beneficial for predicting the masked frames.
	Furthermore, an additional bottleneck structure $L$ is set to perform dimension reduction, which consists of a Multiple Layer Perceptron~(MLP) between $E$ and $D$.
	
	Following conventional AE-based approaches~\cite{Koizumi2020, suefusa2020anomalous}, Mean Squared Error~(MSE) is adopted as an objective,
	\begin{equation} \label{eq_1}
		J = 
		\left|\left|M \odot \x - M \odot D\left(L\left(E\left(\x_m\right)\right)\right)\right|\right|^2_2
	\end{equation}
	where $M$ is a binary mask matrix representing the position of randomly masked frames and $\odot$ is an element-wise multiplication.
	The anomaly score is calculated using the MSE between predicted and target frames.
	
	PAE is pre-trained using the normal sounds from each machine type in an end-to-end manner.
	In GeCo, PAE is further applied for the classification based method, where reconstructed output is used with the original normal data together, under the multi-task learning framework.
	
	\subsection{Joint Generative-Contrastive Learning~(GeCo)}
	The GeCo follows the unified self-supervised representation learning method of~\cite{chen2022self}, where sounds from different machine IDs are used as pseudo anomalies.
	Let $\tilde{\x}$ denote the reconstruction from original normal sample $\x$, and $f_{\theta}$ be the feature extractor parameterized with $\theta$.
	Given a batch of $N$ samples, $\B=\{\x_i, \y_i\}, {i=1,\cdots,N}$, the corresponding reconstructions $\tilde{\B}=\{\tilde{\x}_i, \y_i\}, {i=1,\cdots,N}$, where $\y_i$ is machine ID.
	The minibatch for training GeCo is $\B \cup \tilde{\B}$.
	In operation, only the masked frames in $\x$ are reconstructed by their neighbourhood.
	Sample $\tilde{\x}$ can also be considered as a type of data augmentation from $\x$, and shares the same machine ID as $\x$. 
	Learning from both $\tilde{\x}$ and $\x$ aims to improve the generalization and robustness of the representation. 
	
	From the perspective of contrastive learning~\cite{hojjati2022self}, discriminating $\tilde{\x}$ and $\x$ may enforce the learned representation to be compact.
	We further train GeCo under the multi-task learning framework, as shown in Fig.~\ref{fig1}.
	Given the original $\x$ and reconstruction $\tilde{\x}$, the Cross-Entropy~(CE) loss $\mathcal{L}_{\textit{CE}}$ is utilized, which results in machine ID related clusters.
	The supervised contrastive learning~\cite{khosla2020supervised} then separates the reconstruction $\tilde{\x}$ and original $\x$ points within each class.
	Specifically, for a semantic cluster related to the $c$-th class in each mini-batch, given embedding $\z_i=f_{\theta}(\x_i|\y_i=c)$, the positive samples are $C_p=\{\z_p, p\neq{i}|\y_p=c\}$, and the negative samples are $C_n=\{\tilde{\z}_i=f_{\theta}(\tilde{\x}_i|\y_i=c)\}$.
	The supervised contrastive loss is defined as follows,
	\begin{equation} \label{eq_2}
		\mathcal{L}_{\textit{Con}} = -\frac{1}{|C_p|} \sum_{\z_p \in C_p}
		\log \frac{\exp(\z_i \cdot \z_p)} {\sum_{\z_a \in {C_p \cup C_n} } \exp (\z_i \cdot \z_a)}
	\end{equation}
	where $\left| C_p\right|$ is the cardinality of positive samples.
	By replacing $\z_p$ and $\tilde{\z}_i$ with the corresponding weight vectors $\w_p$ and $\w_n$, we can express the contrastive loss in Eqn.\eqref{eq_2} as a Binary Cross-Entropy~(BCE) loss, as discussed in~\cite{sun2020circle}.
	The total loss in GeCo for multi-task learning is the weighted sum,
	\begin{equation}\label{eq_3}
		\mathcal{L}_{\textit{total}}=\mathcal{L}_{\textit{CE}} + \lambda\mathcal{L}_{\textit{Con}}
	\end{equation}
	where $\lambda$ is the balance factor in total loss.
	When $\lambda$ is set to 0, using PAE to generate samples can be considered a kind of data augmentation.
	
	To further balance the CE and BCE loss, a ramp-up strategy is proposed for adjusting $\lambda$.
	During the early stages of training, $\lambda$ is set to 0, to enable the model to focus on learning robust and generalizable embeddings relevant to the machine IDs.
	Due to the CE loss, the embedding space will begin to contain ID-related clusters.
	$\lambda$ is then linearly increased from 0 to $\lambda_{\textit{max}}$, as the training proceeds.
	This enables the model to gradually pay attention to classifying the reconstruction and original data, without influencing different machine ID classes.
	With the help of this ramp-up strategy, the model can not only maintain its discriminative capabilities according to machine IDs, but also learn a compact representation that is beneficial for the ASD task.
	
\subsection{Anomaly Score Fusion}
	As aforementioned, GeCo integrates generative and contrastive representation learning.
	The former exploits frame-level prediction error as an anomaly score, whereas the latter utilizes clip-level embeddings to calculate a similarity with the ID-related cluster centres.
	A weighted fusion is then computed of frame-level and clip-level scores:
	\begin{equation} \label{eq_4}
		{\rm AnomalyScore}\left(\x\right) ={\rm MSE} + \gamma \left(1- {\rm Cos\mbox{-}Simi}\right) 
	\end{equation}
	where $\gamma$ is the weight hyper-parameter.
	${\rm MSE}$ represents the frame-level anomaly score, as described in section~\ref{section2-1} in detail.
	${\rm Cos\mbox{-}Simi}$ is the cosine-similarity between the embedding of a test clip and the centre of the machine ID clusters, which is applied to calculate the clip-level anomaly score.
	This simple score fusion can effectively make use of the complementarity of the frame-level and clip-level scores to improve overall ASD performance.

\section{Experiments and results}
\label{sec:exp}
\subsection{Dataset}
	We experimentally evaluate GeCo on the DCASE2020 Challenge Task2 development dataset, which is composed of a subset of ToyADMOS~\cite{Koizumi_WASPAA2019_01} and MIMII~\cite{Purohit_DCASE2019_01} datasets.
	The development dataset has six machine types~(ToyCar, ToyConveyor, Fan, Pump, Slider, and Valve). 
	Each machine type consists of clips from three or four machines.
	The real normal sounds from different machines are labeled with machine IDs, leading to a training set having 23 classes in total.

\subsection{Implementation Details}
\begin{table*}[t]
	\setlength\tabcolsep{2.75pt}
	\centering
	\caption{AUC~(\%) and pAUC~(\%) for each machine type obtained from different methods.
The final column averages the AUC and pAUC percentage scores for all machine types. Top-1 refers to the winning system for DCASE2020 task2.}
	\label{table1}
	\renewcommand\arraystretch{0.9}
	\begin{tabular}{lcccccccccccc|cc}
		\hline
		\multicolumn{1}{c}{\multirow{2}{*}{Methods}} & \multicolumn{2}{c}{ToyCar}      & \multicolumn{2}{c}{ToyConveyor} & \multicolumn{2}{c}{Fan}         & \multicolumn{2}{c}{Pump}        & \multicolumn{2}{c}{Slider}      & \multicolumn{2}{c|}{Valve}       & \multicolumn{2}{c}{Average}     \\ \cline{2-15} 
		\multicolumn{1}{c}{}                         & AUC            & pAUC           & AUC            & pAUC           & AUC            & pAUC           & AUC            & pAUC           & AUC            & pAUC           & AUC            & pAUC           & AUC            & pAUC           \\ \hline
		\multicolumn{15}{l}{\textit{Generative Methods}}                                                                                                                                                                                                                                           \\ \hline
		AE~\cite{Koizumi2020}                                           & \textbf{80.90} & 69.90          & 73.40          & 61.10          & 66.20          & 53.20          & 72.90          & 60.30          & 85.50          & 67.80          & 66.30          & 51.20          & 74.20          & 60.58          \\
		IDNN~\cite{suefusa2020anomalous}                                         & 80.19          & \textbf{71.87} & 75.74          & 61.26          & 69.15          & 53.53          & 74.06          & 61.26          & 88.32          & 69.07          & 88.31          & 65.67          & 79.30          & 63.78          \\
		ANP~\cite{wichern2021anomalous}                                          & 72.50          & 67.30          & 67.00          & 54.50          & 69.20          & \textbf{54.40} & 72.80          & 61.80          & 90.70          & 74.20          & 86.90          & 70.70          & 76.52          & 63.82          \\
		PAE                                          & 75.35          & 69.70          & \textbf{77.58} & \textbf{61.37} & \textbf{72.94} & 54.37          & \textbf{74.27} & \textbf{62.01} & \textbf{91.92} & \textbf{74.39} & \textbf{95.41} & \textbf{81.24} & \textbf{81.25} & \textbf{67.18} \\ \hline
		\multicolumn{15}{l}{\textit{Discriminative   Methods}}                                                                                                                                                                                                                                     \\ \hline
		MobileNetV2~\cite{giri2020self}                                  & 87.66          & 85.92          & 69.71          & 56.43          & 80.19          & 74.40          & 82.53          & 76.50          & 95.27          & 85.22          & 88.65          & 87.98          & 84.00          & 77.74          \\
		STgram~\cite{liu2022anomalous}                                       & 88.80          & 87.38          & 72.93          & 63.62          & 91.30 & \textbf{86.73} & 91.25          & 81.69          & \textbf{99.36} & \textbf{96.84} & 94.44          & 91.58          & 89.68          & 84.64          \\
		Baseline                                     & 94.31          & 86.61          & 70.33          & 60.62          & 89.62          & 83.81          & \textbf{93.24} & 84.72          & 97.30          & 92.40          & 97.83 & 91.61          & 90.44          & 83.30          \\
		GeCo                                         & \textbf{96.62} & \textbf{89.33} & \textbf{74.69} & \textbf{65.82} & \textbf{92.73}         & 85.19          & 93.09          & \textbf{86.89} & 98.61         & 95.26          & \textbf{99.06}          & \textbf{95.52} & \textbf{92.47} & \textbf{86.34} \\ \hline
		\multicolumn{15}{l}{\textit{Ensembles}}                                                                                                                                                                                                                                                     \\ \hline
		Top-1~\cite{giri2020self}                                         & 95.57          & 91.54 & \textbf{81.46}          & 66.62          & 82.39          & 78.23          & 87.64          & 82.37          & 97.28          & 88.03          & 98.46          & 94.87          & 90.47          & 83.61          \\
		PAE+GeCo~(fixed-$\gamma$)                                    & 97.17 & 90.98          & 80.91 & \textbf{67.75} & 92.33 & 84.10 & 94.08 & 86.45 & 99.14 & 96.08 & 99.42 & 96.96 & 93.84 & 87.05 \\ 
		PAE+GeCo~(multi-$\gamma$)                                    & \textbf{97.22} & \textbf{91.57}          & 81.01 & 67.63 & \textbf{92.77} & \textbf{84.83} & \textbf{94.10} & \textbf{86.54} & \textbf{99.23} & \textbf{96.26} & \textbf{99.47} & \textbf{97.21} & \textbf{93.97} & \textbf{87.34} \\ \hline
	\end{tabular}
\end{table*}
	We extract the log-Mel spectrogram as the input of our system with 128 Mel filters, a window size of 1024 and hop size of 512, for all 16kHz sample rate input clips.
	 
	PAE is first used to model the input of 5 frames for every machine type. To do this, a random frame is selected as the target frame and is replaced with a learnable mask token.
	Both encoder $E$ and decoder $D$ consist of two transformer blocks having dimensions of 512 and 256, respectively, with the dimension of the sandwiched bottleneck structure $L$ being 64.
	PAE is trained using the ADAM optimizer~\cite{kingma2014adam} over 60 epochs with a batchsize set to 512.
	An initial learning rate of 1e-3 is used for training during the first 30 epochs, declining to 1e-4 for the remaining 30 epochs.

	We utilize all of the training data in the development dataset to train GeCo.
	65 frames are randomly cropped as the input, and then one random frame out of every 5 frames within the input features is substituted by the output of PAE.
	The ResNet18~\cite{chen2022self} operates as a feature extractor, with a mini-batch size of 32, which would expand to 64 after generation of the PAE output.
	A SGD algorithm is employed with momentum of 0.9 and weight decay of 1e-4 to optimize this model over 120 epochs, with an initial learning rate of 0.1. This is multiplied by 0.1 at the 50th and 90th epochs.
	As for ramp-up strategy, the value of $\lambda$ is set to 0 in the first 30 epochs and gradually increases to 10 from the 30th epoch to the 90th epoch.
	$\lambda$ stays at $\lambda_{\textit{max}}=10$ for the last 30 epochs. 
	
	Performance is evaluated by using the area under the receiver operating characteristic curve~(AUC) and the partial-AUC~(pAUC), which is calculated as the AUC over a low false-positive-rate~(FPR) range which is set to $\left[0, 0.1\right]$ for DCASE2020 Challenge Task2.
	
\subsection{Results}
	We compare the proposed methods with previous generative and discriminative architectures and present the performance metrics obtained for each machine type in Table~\ref{table1}.
	In discriminative methods, the baseline system is a vanilla ResNet18~\cite{chen2022self} that uses all training data to recognize 23 machine IDs.
	Besides, we compare our anomaly score fusion results with the highest ranked ensemble systems in the DCASE2020 Challenge Task2.
	
	\begin{table}[b]
		\centering
		\caption{Average AUC~(\%) and pAUC~(\%) results of ablation experiments with various balance factors $\lambda$.}
		\label{table2}
		  \renewcommand\arraystretch{0.9}
		\newcolumntype{L}[1]{>{\raggedright\let\newline\\\arraybackslash\hspace{0pt}}m{#1}}
		\newcolumntype{C}[1]{>{\centering\let\newline\\\arraybackslash\hspace{0pt}}m{#1}}
		\begin{tabular}{L{2.5cm}C{2cm}C{1cm}}
			\hline
			System            & AUC            & pAUC           \\ \hline
			Baseline              & 90.44          & 83.30          \\
			$\lambda$=0           & 91.50          & 84.63          \\
			$\lambda$=1           & 91.88          & 85.20          \\
			$\lambda$=10          & 90.09          & 82.65          \\
			with $\lambda$ ramp-up               & \textbf{92.47} & \textbf{86.34}         \\ \hline
		\end{tabular}
	\end{table}
	For generative methods, the performance of PAE is superior to other generative models for most machine types, especially the Valve, proving the validity of exploiting context information for ASD.
	For discriminative methods, it is evident that GeCo achieves significant improvement over the baseline and other methods.
	We believe the primary reason is that the representations learned by our proposed method can more effectively describe a compact distribution of normal samples.
	Meanwhile, due to the constraint of $\mathcal{L}_{\textit{CE}}$ and $\mathcal{L}_{\textit{Con}}$, slight anomalous discrepancies can be reflected in the feature representations, which enables hard anomalies to be detected more efficiently, resulting in a higher pAUC.
	Employing anomaly score fusion~(`Ensembles'), we fix $\gamma$ in Eqn~\eqref{eq_4} to 200 and present the results in Table~\ref{table1}.
	In addition, we achieve the best performance by grid searching the optimal value of $\gamma$ for each machine type.
	The optimal values are presented as \textit{machine type($\gamma$)} as follows: ToyCar(125), ToyConveyor(135), Fan(495), Pump(225), Slider(110), Valve(125).
	It is not hard to observe that our method significantly outperforms the state-of-the-art ensemble method~\cite{giri2020self}.
	We believe this is due to the extraordinary fact that our method introduces a generative model into a discriminative framework, which allows frame-level and clip-level anomaly scores to be calculated together.
	Moreover, we also validate the effectiveness of our method on the evaluation set.
	We achieve 96.67$\%$ AUC and 90.61$\%$ pAUC, which outperforms the 96.42$\%$ AUC and 89.24$\%$ pAUC achieved by the state-of-the-art system\cite{9534290}.
	
	In Table~\ref{table2}, we study the influences of the balance factor $\lambda$ on performance.
	When $\lambda=0$, the reconstruction samples effectively act as augmented samples for training the model, which achieves noticeable improvement over the baseline.
	This result indicates reconstruction samples are robust data augmentation, which is consistent with~\cite{xu2022masked}.
	However, if the value of $\lambda$ is unreasonable~(\textit{e.g.} $\lambda=10$), the result performs worse than the baseline.
	That is because the model focuses more on detecting generative frames, resulting in loose machine ID related clusters and imprecise representations of normal samples.
	The ramp-up strategy yields the best performance, suggesting its effectiveness in solving the sensitivity problem towards $\lambda$.

\section{conclusion}
\label{sec:conclusion}
	In this paper, we propose a joint GeCo method for ASD tasks.
	The method includes a PAE with self-attention mechanism which is exploited to perform frame-level prediction. 
	With the help of a self-attention mechanism, context information from neighbouring frames to the masked ones is effectively captured to improve ASD performance, based on generative learning methods.
	The output of the PAE plays several crucial roles under a multi-task framework. 
	As augmented samples, the reconstructions improve the generalization capability and robustness of the representation obtained by self-supervised learning.
	Meanwhile, as negative samples from supervised contrastive learning, they act to make the embedding space of normal data more compact.
	Extensive experimental results demonstrate the superiority of our proposed methods, and we achieve state-of-the-art performance on the evaluation set.

\vfill\pagebreak

\bibliographystyle{IEEEbib}
\bibliography{refs}

\end{document}